\documentclass[fleqn,usenatbib]{mnras}

\usepackage{newtxtext,newtxmath}

\usepackage[T1]{fontenc}
\usepackage{ae,aecompl}

\usepackage{graphicx}	
\usepackage{amsmath}	

\usepackage[dvipsnames]{xcolor}

\title[Gaseous DF]{Gaseous Dynamical Friction: a Numerical Study of Extended Perturbers}

\author[B. Morton et al.]{
B. Morton,$^{1}$\thanks{E-mail: morton@roe.ac.uk}
S. Khochfar,$^{1}$
J. O$\tilde{\textrm{n}}$orbe,$^{2}$
\\
$^{1}$Institute for Astronomy, Royal Observatory, Edinburgh EH9 3HJ, UK\\
$^{2}$Facultad de Fisica, Universidad de Sevilla, Sevilla, Spain
}

\date{Accepted XXX. Received YYY; in original form ZZZ}

\pubyear{2021}

\begin{document}
\label{firstpage}
\pagerange{\pageref{firstpage}--\pageref{lastpage}}
\maketitle

\begin{abstract}
The process of momentum and energy transfer between a massive body and a background medium it is moving through is known as dynamical friction (DF). It is key to our understanding of many astrophysical systems. We present a series of high-resolution simulations of gaseous DF using Lagrangian meshless finite mass hydrodynamics solver, the moving-mesh MUSCL scheme, and the piecewise parabolic method (PPM) solver. We use a set of simulations of massive bodies, modelled as Plummer spheres, moving with Mach $0.2 \leq \mathcal{M} \leq 3$. We investigate at which radial distances from the perturber these solvers recover the linear point mass solution for gaseous DF. We analyse the drag force and the structure and time evolution of the wake. The different solvers agree closely. Numerical convergence is reached when the initial spatial resolution is $0.2r_s$, where $r_s$ is the softening scale of the Plummer sphere. We find that the wake structure and drag force are recovered, at the $5\%$ level, when compared beyond $4r_\mathrm{s}$. Our results predict that models using the standard linear point mass DF solution will overestimate the drag force on extended perturbers by as much as 25\%, for Mach$\sim$1. Finally, we consider DF in the context of galaxy clusters, where dark matter subhaloes move through circumgalactic media. We show that DF is typically in the linear regime for most subhaloes in hosting haloes $<10^{11}$ M$_{\odot}$ but non-linear in more massive host haloes.
\end{abstract}




\section{Introduction} \label{sec:intro}

As a massive object moves through a background medium, gravitational interactions build an over-dense wake of material behind the massive traveller. Momentum and energy are transferred from this object, referred to as the perturber, to the surrounding medium. This process is known as dynamical friction (DF) and is key to our understanding of the evolution of a number of astronomical systems. These systems cover very different mass and length scales. DF drives processes from galaxy clusters \citep{ja:elzant2004, ja:kim2005, ja:adhikari2016}, to satellite galaxies orbiting within their host halo \citep{ja:zhao2004, ja:fujii2006, ja:ogiya2016}, supermassive black hole formation \citep{ja:beckmann2017}, compact object binaries and mergers \citep{ja:just2010, ja:dosopoulou2017, ja:tagawa2018}, galaxy mergers \citep{ja:khochfar2001}, and planets within disks \citep{ja:teyssandier2012}. In the case of galaxy clusters, dark matter (DM) subhaloes experience DF as they perturb the circumgalactic medium (CGM) of their host halo. The exact number, and spatial distribution, of these DM substructures varies with different cosmological models. Large scale cosmological simulations are a key tool in testing these models, and the comparison of simulated and observed DM substructure is currently a key challenge for $\Lambda$CDM \citep{ja:weisz2019}.

As an example, in the case of galaxy clusters, DF plays a key role in driving the accretion of substructure. Hierarchical collapse, within the standard $\Lambda$CDM cosmological model, predicts the presence of large numbers of low mass subhaloes within higher mass hosts. The number density of satellites found in simulations follows a power law, with a slope of approximately $-2$, predicting a steep increase in the number density of satellites going to lower masses \citep{ja:springel2008}. These structures will experience both \textit{collisionless} and \textit{collisional}, or gaseous, DF. These emerge as the substructure moves within the extended DM and CGM of the host. DF provides a mechanism by which the substructure sheds angular momentum and energy, allowing it to merge with the central galaxy \citep{ja:boylankolchin2008, ja:khochfar2008}. In this way, DF drives the build-up of structure in a cold dark matter (CDM) cosmology. These mergers are a key driver of the evolution of galaxies, providing fuel for star formation and disrupting galactic structures \citep{ja:barnes1996, ja:beckman2008, ja:robaina2010, ja:khochfar2011, ja:somerville2015}. Gaseous DF is particularly pronounced at high redshifts, where halo gas-mass fractions are highest \citep{ja:daddi2010, ja:tacconi2010}.

The structure of the overdense wake can be separated into two distinct contributions, one from a collisionless medium, such as background stars or cold dark matter, and one from a collisional component, typically baryonic gas. The additional pressure forces present in collisional media result in significant differences in the resultant wake, and in the drag force, when compared to the collisionless case \citep{ja:ostriker1999}. This is most pronounced for Mach numbers close to $\mathcal{M}=1$, where Mach has the standard definition as the ratio of bulk velocity of the gas $v$ to its sound speed $c_s$, $\mathcal{M}=v/c_s$. We use the standard definitions for subsonic $v<c_s$, $\mathcal{M}<1$, and super-sonic $v>c_s$, $\mathcal{M}>1$, with the transition point at $v=c_s$, $\mathcal{M}=1$. This assumes that the wake remains in the linear regime. \footnote{Here the linear regime refers to wakes where the density is less than double the initial value, at a given radius.} As the induced wake becomes increasingly non-linear, the gaseous DF drag force decreases \citep{ja:kim2009}. The complex structure of gravitationally induced wake has previously been studied using both analytic \citep{ja:just1990, ja:ostriker1999, ja:namouni2010} and numerical approaches \citep{ja:sanchez1999, ja:sanchez2001, ja:kim2007, ja:kim2009, ja:lee2011, ja:thun2016, ja:park2017, ja:tomoya2024, ja:bernal2013}. With a simplified, adiabatic, single-phase, ideal-gas background, the growth of the wake is effectively scale-free, allowing the analytic predictions to be applied across scales.

The analytic collisionless solution is found by calculating a sum of two body interactions, integrated over all time, between the perturber and an infinite background medium \citep{ja:chandrasekhar1943}. The analytical solution to the collisional case can be obtained in the linear regime, producing an approximation of the time dependent structure of the wake using linear perturbation theory \citep{ja:just1990, ja:ostriker1999}. Before this only the steady state analytic solution had been studied \citep{ja:bondi1944}. Within the linear approximation of the analytic treatment, the collisional drag force is significantly higher for perturbers moving with Mach numbers $0.7 < \mathcal{M} < 2$.

A similar trend is observed in numerical simulations for low-mass, extended, perturbers \citep{ja:sanchez1999, ja:kim2009, ja:bernal2013}. These studies modelled perturbers as effectively collisionless Plummer  spheres with some gravitational softening $r_s$. Here, gravitational softening refers to the truncation of the potential at small radii, introduced by adding this softening scale $r_s$ into the radius term of the potential. Such a potential no longer diverges as $r\rightarrow0$. This potential can be seen as distributing the mass of the object, instead of modelling a true point mass. These studies did not model either accretion on to the perturber, or face-on collisions with the gaseous medium. This is why it is effectively collisionless. Importantly, the analytic point mass prediction for the wake's structure is only accurate in the linear to slightly non-linear regime. 



More complex scenarios, such as those that include more physical processes, have also been studied, such as rigid perturbers that experience both gravitational drag, and the drag from face-on collisions with the surrounding medium \citep{ja:thun2016}, as well as perturbers on orbital trajectories \citep{ja:sanchez2001, ja:kim2007}, and the effect of radiative feedback from an accreting black hole on the drag force \citep{ja:toyouchi2020}. The analytic solution for the wakes of objects moving on circular trajectories have also been calculated \citep{ja:dejaques2022, ja:eytan2024}, as has the solution for accreting objects \citep{ja:lee2011}.


In the last decade, a new hybrid class of methods has been developed, utilising moving unstructured mesh approaches that attempt to merge the advantages of Eulerian grid-based and Lagrangian particle-based methods. They aim to provide the instability and shock capture efficiency of grid-based methods with the natural resolution adaptation of particle methods. Within astrophysics, these methods are still relatively new, although several established codes have emerged \citep{ja:springel2009, ja:duffell2011, ja:hopkins2015, ja:duffell2016, ja:springel2020}. These have been shown to be generally competitive with the established multi-physics simulation codes \citep{ja:wadsley2004, ja:springel2005, ja:bryan2014, ja:gonnet2014, ja:menon2014, ja:wadsley2017}. They are ideal for studying the effects of DF, as their natural resolution and stability to shocks should allow the capture of the sharp cone of the supersonic wakes. However, idealised DF studies have not been attempted with modern moving-mesh methods.

The impact of DF is a topic of significant ongoing study in many different scenarios. The point mass drag force prescription from \cite{ja:ostriker1999} is frequently utilised to model the effects of gaseous dynamical friction, in both semi-analytic models and as a subgrid model in more detailed simulations, including those of the trajectories and mergers of stars \citep{ja:tagawa2018, ja:antoni2019, ja:delaurentiis2023}, of black holes \citep{ja:pfister2019, ja:volonteri2020, ja:ma2021, ja:chen2022}, the circularization of orbits within discs \citep{ja:bonetti2020}, the tidal stripping of galaxies \citep{ja:jackson2010} and the cool flows in galaxy clusters \citep{ja:kim2005}. Comparing numerical results of extended perturbers to the analytic point mass predictions will allow to estimate the error on DF force calculations in such models. As noted above, this work is applicable to situations where face-on collisions and accretion do not play a significant role in the dynamical evolution of the perturbing object. The effective resolution available to capture the gravitationally induced wake self-consistently, even in the most cutting-edge simulations, is often low. In other words, a relatively small number of particles, or cells, resolve the mass and length scales of the wake. To arrive at robust conclusions from cosmological simulations of galaxy formation, it is therefore necessary to understand the impact very low mass and length resolutions might have on the formation and evolution of the dynamical friction wake.

In this work, we present a set of idealised, three-dimensional gravo-hydrodynamic simulations of massive Plummer spheres, embedded within a uniform region of adiabatic gas moving with some velocity. In Section \ref{sec:analytic} we summarise the relevant analytic predictions. Section \ref{sec:num_method} describes the numerical setup that we use, and \ref{sec:results} contains our results, showing the predicted wake, the force across a range of Mach numbers, and the impact of resolution. In Section \ref{sec:discussion} we discuss these results, and examine the implication of our findings for dynamical friction experienced by dark-matter subhaloes in a cosmological context.

\section{Analytic solution to dynamical friction in gaseous media} \label{sec:analytic}

In contrast to dynamical friction in a collision-less medium, pressure gradients within a gaseous medium introduce additional forces on the evolving gas. These forces lead to the formation of structure within the overdense wake, where the wake is defined as the perturbation to the uniform density background gas. The time-dependent form of this structure can be derived using linear perturbation theory to model the density perturbation as an expanding density wave \citep{ja:ostriker1999}.

\subsection{Linear Wake} \label{sec:ana_lin}
Using linear perturbation theory, \citet{ja:ostriker1999}\footnote{See \citet{ja:just1990} for an alternative derivation.} produces an analytic prediction for the force of DF felt by a point mass perturber $M_{\mathrm{p}}$, moving at constant velocity $V_0$ in the $z$-direction through an adiabatic gaseous medium with infinite uniform density $\rho_0$, where the gas does not have self-gravity. The medium has a far-field sound speed $c_s$, and the perturber moves with Mach number $\mathcal{M} \equiv V_0/c_s$. The resultant prediction for the density distribution of the induced wake, $\rho = \rho_0(1+\alpha)$, described by the over-density $\alpha(s,R,t)$, is
\begin{equation} \label{eq:alpha}
    \alpha=\frac{GM_{\mathrm{p}}/c_s^2}{[s^2+R^2(1-\mathcal{M}^2)]^{1/2}} \\
    \times
    \begin{cases}
         1 & \parbox{3.0cm}{\textrm{if} \hspace{0.5mm} $R^2+z^2<(c_s t)^2$} \\
         2 & \parbox{3.0cm}{\textrm{if} \hspace{0.5mm} $\mathcal{M}>1, \\ R^2+z^2>(c_s t)^2, \\  s/R<-(\mathcal{M}^2-1)^{1/2}, \\ z>c_s t/\mathcal{M}$} \\
         0 & \textrm{otherwise}
    \end{cases}
\end{equation}
where $s=z-V_0t$ is the distance from the perturber along the direction of travel and $R$ is the cylindrical radius. The $R$ coordinate  gives the radial distance from the axis aligned with the initial motion. It is analogous to the impact parameter in the collisionless case. In this frame, the gas effectively moves past the perturber, with the initial velocity in the negative $s$-direction. This is equivalent to having a moving perturber travelling through a static medium. It should be noted that $s=0$ is the position of the perturber at all times. This can be seen in Figure \ref{fig:wake_zoom}, with the peak in the wake always appearing at $s=0$. The coordinate $s$ is effectively just the $z$ position with the origin moved to the position of the perturber at time $t$. Formally, this distribution diverges as the spherical radius $r\rightarrow0$. However, the derivation assumes linearity, so the solution is only valid in the parts of the wake that remain linear. Ostriker therefore states this wake solution is only valid for this point-mass at $r \gtrsim r_{\mathrm{min}}$, where $r_\mathrm{min}$ must be chosen such that the wake outside this radius it is linear, and with

\begin{equation} \label{eq:wake_limits}
    \frac{GM_{\mathrm{p}}}{c_s^2} \lesssim r_{\mathrm{min}} \lesssim (V_0-c_s)t.
\end{equation}
The lower limit ensures that all parts of the wake included in this analysis are within the linear regime, while the upper limit requires that the wake extends beyond $r_\mathrm{min}$ itself. The lower limit is also effectively the Hoyle-Lyttleton radius \citep{ja:hoyle1939}. This, and its successor the Bondi-Hoyle-Lyttleton radius \citep{ja:bondi1952}, which replaces sound speed with gas bulk velocity, to account for relative motion, estimate the radius within which material will be accreted onto the perturbing object. As accretion is not included, this is a natural restriction to use, particularly as it also emerges from the analytic calculation as this limit on the linear region.

\begin{figure}
    \centering
    \includegraphics[width=0.45\textwidth]{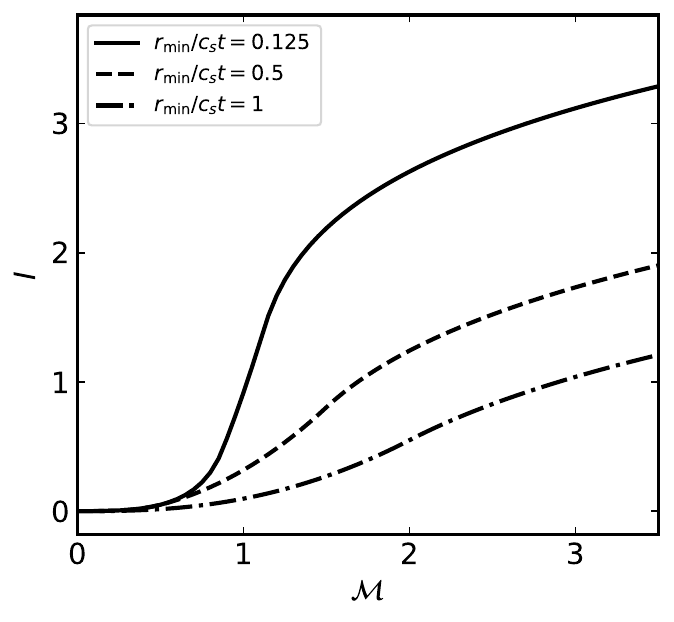}
    \caption{Value of integral $I$ as a function of Mach number $\mathcal{M}$, for $r_{\mathrm{min}}/c_s t = 0.125, 0.5, 1$. Over time, the perturbation in density grows across Mach number, as we move from the lower to the upper curve.}
    \label{fig:I_mach}
\end{figure}

Still following \citet{ja:ostriker1999}, and by integrating over this overdense wake from $r_{\mathrm{min}}$ to the edge of the wake, using
\begin{equation} \label{eq:force_integral}
    F_{\rm{DF}} = 2\pi GM_{\mathrm{p}}\rho_0\iint R\frac{\alpha(t)s}{(s^2+R^2)^{3/2}}dsdR,
\end{equation}
the drag force created by new mass distribution asymmetry takes the form
\begin{equation} \label{eq:force}
    F_{\mathrm{DF}}=-\frac{4\pi(GM_{\mathrm{p}})^2\rho_0}{V_0^2}I.
\end{equation}
The parameter $I$ represents the integral over the over-density, divided by the collisionless solution (see Equation (13) of \citet{ja:ostriker1999} for exact form). It is the factor by which the DF force differs from the purely collisionless case. It depends on Mach number, with
\begin{equation} \label{eq:isub}
    I_{\mathrm{sub}}=\frac{1}{2}\ln{\left(\frac{1+\mathcal{M}}{1-\mathcal{M}}\right)}-\mathcal{M}
\end{equation}
for the subsonic perturbers, and
\begin{equation} \label{eq:isuper}
    I_{\mathrm{super}}=\frac{1}{2}\ln{\left(\frac{\mathcal{M}+1}{\mathcal{M}-1}\right)}+\ln{\left(\frac{\mathcal{M}-1}{r_{\mathrm{min}}/c_s t}\right)}
\end{equation}
for the supersonic perturbers. The parameter $I$ as a function of the Mach number is shown in Figure \ref{fig:I_mach}. This result is only accurate when all regions of the wake included in the force calculation remain linear. In other words, this solution assumes $\alpha\lesssim1$. The force predicted by this approach is greater than the equivalent collisionless force, for a range of Mach numbers $0 \lesssim \mathcal{M} \lesssim 3$ (see Figure 3 of \cite{ja:ostriker1999}). The largest difference comes when we are close to $\mathcal{M}=1$, where the collisional prediction is undefined. It should be noted that the collisional drag force deviates by a factor of a few from the pure collisionless case. This increases with time, as the collisional solution is time dependent, while the collisionless is not. The $\ln{(c_st/r_\mathrm{min}})$ part of the supersonic solution is the analogue to the Coulomb logarithm $\ln{(b_\mathrm{max}/b_\mathrm{min})}$ in the collisionless case \citep{ja:chandrasekhar1943}.

This prediction has been tested against numerical solutions of a softened potential \citep{ja:kim2009, ja:bernal2013}, and we follow this approach to allow comparison to earlier work. The potential $\Phi(r)$, which defines the perturber, is typically that for a Plummer sphere \citep{ja:plummer1911}, with
\begin{equation}
    \Phi(r) = - \frac{GM_{\mathrm{p}}}{\left(r^2+r_\mathrm{s}^2\right)^\frac{1}{2}},
\end{equation}
with mass $M_{\mathrm{p}}$ and softening scale $r_\mathrm{s}$. This is equivalent to modelling a perturber with an extended mass distribution, but still a collisionless one, which is most closely related to dark matter subhaloes or clusters of stars. The softening is analogous to the extent of the perturbing object, and is used to characterise the time scales of the simulation via the sound speed crossing time of this length $t_{c}=r_\mathrm{s}/c_\mathrm{s}$.

The condition for the wake to remain linear, i.e. $\alpha<1$, can be parameterised using the dimensionless parameter $A$. For 
\begin{equation} \label{eq:a}
    A = \frac{GM_{\mathrm{p}}}{c_s^2 r_s},
\end{equation}
we require that $A<1$ for the linear assumption to strictly hold. This estimates the density perturbation at radius $r_s$. As mentioned above, this relationship of radius, mass and gas sound speed is reminiscent of estimates of the radius within which material will be accreted. By restricting ourselves to regions where $\alpha<1$ we exclude parts of the wake where accretion would likely play a significant role.


If one uses a true point mass in such a numerical study, particles of gas can experience shot noise. This happens because the gas tracer can get infinitesimally close to the position of the point mass. The diverging potential of said point mass then produces an extreme acceleration on the tracer, flinging the gas away at unphysical speeds. It is therefore necessary to use a perturber with a softened potential, to avoid the shot noise and other artifacts introduced by simulating a point mass perturbing a gas with both finite mass and finite spatial resolutions. Using a softened potential changes the shape of the wake in mostly minor ways. Most notably, in the supersonic case, it spreads out the sharp Mach cone front \citep{ja:furlanetto2002, ja:kim2009}.

The $A$ parameter effectively places a constraint on the physical scenarios that can be described by this prediction, since cases with very high mass perturbers, very small radii, or very large sound speeds, will violate the small $A$ condition. For instance, a star moving through the ISM will have a much smaller $A$ number than the same star in a giant molecular cloud, where the sound speed of the gas is higher.

\section{Numerical Method} \label{sec:num_method}

We run a set of idealised 3D gravohydrodynamic simulations of DF, using the multiphysics codes \texttt{Enzo} \citep{ja:bryan2014}, \texttt{Gizmo} \citep{ja:hopkins2015}, \texttt{Gadget-4} \citep{ja:springel2020} and \texttt{Arepo} \citep{ja:springel2009}, comparing their output with both analytic predictions and previous numerical results. In this section, we describe our numerical approximation of the idealised case for DF, the key parameters used in the code setup, and the initial conditions of the simulations.

\subsection{Setup} \label{sec:num_setup}
The numerical setup is designed to closely replicate the assumptions made in the derivation of the point mass solution. The initial conditions have a uniform density $\rho_0$ adiabatic gas, moving with a bulk velocity $V_0$, in a 3D box with a fixed gravitational potential $\Phi$ from a massive perturber $M_{\mathrm{p}}$. The box has periodic boundaries in all directions. As such, the simulation analysis was limited to times before the wake reached the boundaries of the box, and the gravitational component (i.e. the perturber's potential) was excluded from the periodicity. The bulk velocity has the Mach number $\mathcal{M}=V_0/c_s$. The gas does not experience self-gravity.

To reproduce the uniform initial density, we generate glass-like initial conditions (ICs) using WVTICs \citep{ja:donnert2017,ja:arth2019}. This combines a weighted Voronoi mesh with a particle shuffling method to reproduce arbitrary density distributions with a set of SPH particles. The ICs created with this method were found to be superior to simple initial grids and random particle placements, minimising both periodic oscillations in the net force on the perturber and initial variation in the local density distribution. For the \texttt{Arepo} comparison runs, we use these particle positions as the cell centres for the initial mesh.

The idealised set-up is effectively scale free, characterised by $A$ and Mach number $\mathcal{M}$ \citep{ja:kim2009}. The wake grows linearly with sound speed $c_s$, with the deviation from the linear point mass solution dictated by the $A$ parameter. If one normalises the spatial coordinates by $c_st$, the wake should not change shape with time, as long as the limits in Equation (\ref{eq:wake_limits}) are respected. The $A$ parameter sets the relationship between the perturber, from its mass and softening, a proxy for the extent of the perturber, and the medium, through the speed of sound.

The relative motion is captured by the Mach number. We run setups for a range of Mach numbers, exploring the regime where collisional DF differs most from collisionless DF, for $0.7 < \mathcal{M} < 2$. We explore a range of resolutions, dictated by the number of particles. The Mach range is explored with $N=256^3$, $512^3$, $1024^3$ particles, with the most samples at $512^3$. The different simulations are summarised in Table \ref{tab:runs}. The size of the box $L$ is varied in certain runs. Varying the particle number is used to effectively test the convergence of our results with resolution.

To test the numerically modelled gas under different conditions, it is necessary to explore the scale-free parameter space of $A$ and Mach number. The setup is constrained by the periodic boundary conditions, setting an upper limit on how long the scenario can be run before the wake reaches the edge of the box, and so wraps around. This time is defined by the edge length of the box and the initial velocity of the gaseous background. A range of Mach numbers were probed, exploring the regime around the strongest gaseous signal at $\mathcal{M}=1$. The majority of the boxes are run for $t=15t_c$, with the larger boxes pushed to $t=150t_c$, but at the cost of reduced mass resolution, critical to recovering the small over-densities in the wake structure. We have checked that the lowest resolution runs are still retaining acceptable mass resolution.

\begin{table}
    \centering
    \caption{Details of all the simulation runs. Solvers refer to either the \texttt{Gizmo}, \texttt{Gadget4}, \texttt{Arepo}, or \texttt{Enzo} codes, using either the MFM, PPM or MUSCL schemes. $N$ is the number of particles or cells, per edge length, $t_\mathrm{final}$ the end time of the simulation, in units of sound speed crossing time $t_c$ of the softening length $r_s$, and $L$ is the length of the edge of the cube box, relative to our standard box.}
    \label{tab:runs}
    \begin{tabular}{llllll}
        \hline
        ID & Solver & $N$ & $\mathcal{M}$ & $t_\mathrm{final}$ & $L$ \\
        \hline
        GM256M07 & MFM  & $256$ & $0.7$  & 15 & 1\\
        GM256M09 & MFM  & $256$ & $0.9$ & 15 & 1 \\
        GM256M1 & MFM  & $256$ & $1.01$ & 15 & 1 \\
        GM256M11 & MFM & $256$ & $1.1$ & 15 & 1 \\
        GM256M13 & MFM & $256$ & $1.3$ & 15 & 1 \\
        GM2563M15 & MFM & $256$ & $1.5$ & 15 & 1 \\
        GM256M2 & MFM & $256$ & $2$ & 15 & 1 \\
        GM512M02 & MFM & $512$ & $0.2$ & 15 & 1 \\
        GM512M07 & MFM & $512$ & $0.7$ & 15 & 1 \\
        GM512M09 & MFM & $512$ & $0.9$ & 15 & 1 \\
        GM512M1 & MFM & $512$ & $1.01$ & 15 & 1 \\
        GM512M11 & MFM & $512$ & $1.1$ & 15 & 1 \\
        GM512M13 & MFM & $512$ & $1.3$ & 15 & 1 \\
        GM512M15 & MFM & $512$ & $1.5$ & 15 & 1 \\
        GM512M2 & MFM & $512$ & $2$ & 15 & 1 \\
        GM512M3 & MFM & $512$ & $3$ & 15 & 1 \\
        GM1024M07 & MFM & $1024$ & $0.7$ & 15 & 1 \\
        GM1024M09 & MFM & $1024$ & $0.9$ & 15 & 1 \\
        GM1024M1 & MFM & $1024$ & $1.01$ & 15 & 1 \\
        GM1024M11 & MFM & $1024$ & $1.1$ & 15 & 1 \\
        GM1024M13 & MFM & $1024$ & $1.3$ & 15 & 1 \\
        GM1024M15 & MFM & $1024$ & $1.5$ & 15 & 1 \\
        GM1024M2 & MFM & $1024$ & $2$ & 15 & 1 \\
        GM64M13 & MFM & $64$ & $1.3$ & 15 & 1 \\
        GM128M13 & MFM & $128$ & $1.3$ & 15 & 1 \\
        GM256M13 & MFM & $256$ & $1.3$ & 15 & 1 \\
        GM512M13LM & MFM & $512$ & $1.3$ & 15 & 1 \\
        GM1024M13L & MFM & $1024$ & $1.3$ & 150 & 10 \\
        EP512M07 & PPM & $512$ & 0.7 & 15 & 1 \\
        EP512M09 & PPM & $512$ & 0.9 & 15 & 1 \\
        EP512M11 & PPM & $512$ & 1.1 & 15 & 1 \\
        EP512M13 & PPM & $512$ & 1.3 & 15 & 1 \\
        EP512M15 & PPM & $512$ & 1.5 & 15 & 1 \\
        EP512M2 & PPM & $512$ & 2.0 & 15 & 1 \\
        AM512M07 & MUSCL & $512$ & 0.7 & 15 & 1 \\
        AM512M09 & MUSCL & $512$ & 0.9 & 15 & 1 \\
        AM512M1 & MUSCL & $512$ & 1.01 & 15 & 1 \\
        AM512M11 & MUSCL & $512$ & 1.1 & 15 & 1 \\
        AM512M13 & MUSCL & $512$ & 1.3 & 15 & 1 \\
        AM512M15 & MUSCL & $512$ & 1.5 & 15 & 1 \\
        AM512M2 & MUSCL & $512$ & 2.0 & 15 & 1 \\
        AM64M13 & MUSCL & $64$ & 1.3 & 15 & 1 \\
        AM128M13 & MUSCL & $128$ & 1.3 & 15 & 1 \\
        AM256M13 & MUSCL & $256$ & 1.3 & 15 & 1 \\
        \hline
    \end{tabular}
\end{table}

\subsection{Codes}

We primarily asses the gravitationaly induced wake, and resulting effects, from the Lagrangian finite mass (MFM) \citep{ja:hopkins2015} solver in \texttt{Gizmo}. The MFM solver is a hydrodynamics solver that discretises the volume using tracer particles and a chosen kernel function, and then solves the Riemann problem between neighbouring particles, utilising a high order gradient estimator. We use 32 neighbours, and the cubic kernel, for the kernel construction. To augment this assessment, we include comparison runs using the peicewise-parabolic method (PPM) in \texttt{Enzo} \citep{ja:bryan2014} and the moving mesh MUSCL-Hancock implementation in \texttt{Arepo} \citep{ja:springel2009}.


We use the PPM solver in \texttt{Enzo} for comparison to the widely used grid based solver. This implementation is based on a modified version of the PPM method from \citet{ja:colella1984}, adapted for cosmological settings. As the name implies, it solves the Eulerian fluid equations via piecewise parabolic reconstruction of the fluid state across a spatially dicsretised domain. We do not utilise the adaptive mesh refinement capabilities in \texttt{Enzo} in order to maintain the simplest resolution comparison.

Finally, we compare with results from the MUSCL-Hancock moving mesh scheme used in \texttt{Arepo}. This couples the slope-limited piece-wise reconstruction step with spatial decomposition that moves approximately with the fluid, aiming to combine the Lagrangian nature of particle codes with the precision of grid based approaches.

\begin{figure*}
    \centering
    \begin{tabular}{cc}
        \includegraphics[width=0.4\textwidth]{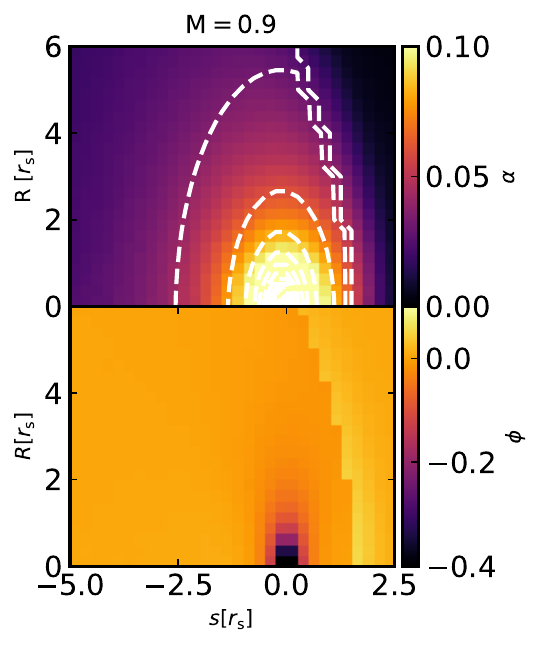} &
        \includegraphics[width=0.4\textwidth]{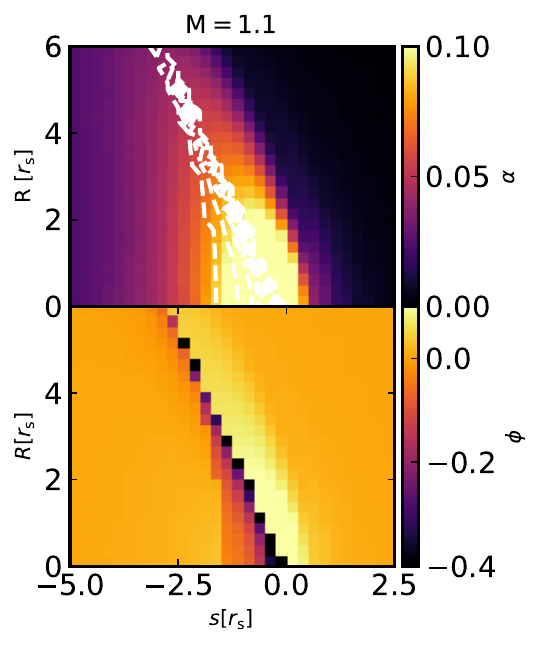} \\
        \includegraphics[width=0.4\textwidth]{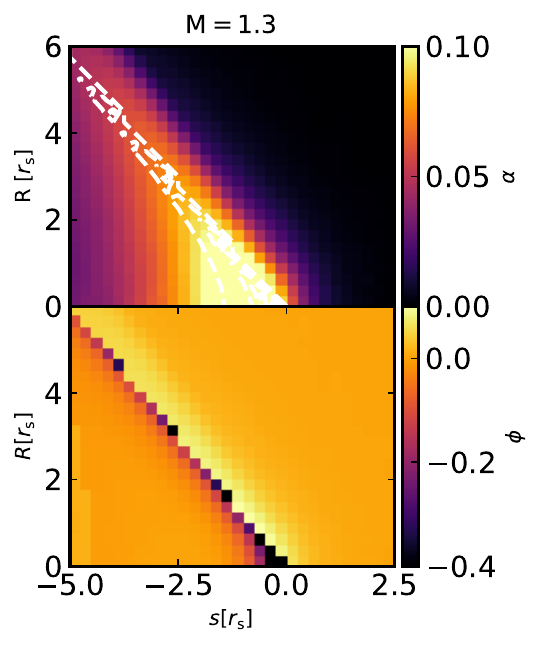} &
        \includegraphics[width=0.4\textwidth]{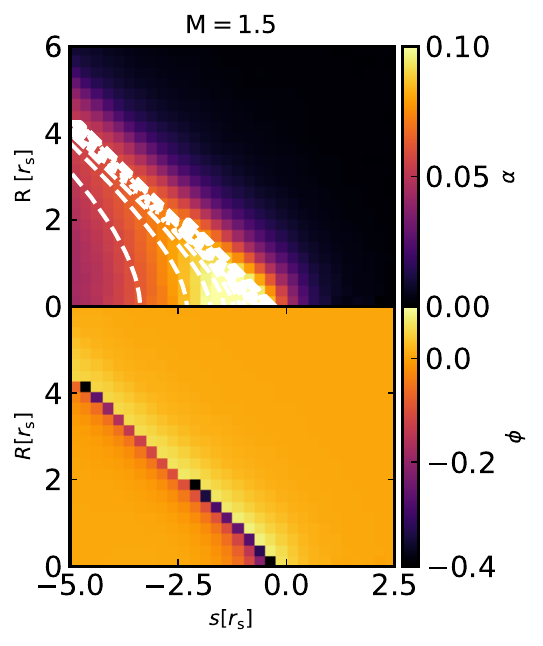}
    \end{tabular}
    \caption{Zoomed view of over-density across $\mathcal{M}=0.9,1.1,1.3,1.5$, at $t=15t_c$, using $N=1024^3$ (runs GM1024M09, GM1024M11, GM1024M13, GM1024M15). The upper part of each panel shows the numerical over-density $\alpha_\mathrm{num}$ in colour and the analytic prediction for the over-density $\alpha_\mathrm{ana}$ as white dashed contours. The contours are provided as a guide to show the shape of the expected shape for a  point mass perturber. We see the development of a bow shock like structure ahead of the perturber. The difference between these distributions $\phi = \alpha_\mathrm{num} - \alpha_\mathrm{ana}$ is shown in the lower part of each panel. The results diverge most dramatically close to the perturber, and along the cone front.}
    \label{fig:wake_zoom}
\end{figure*}

\begin{figure}
    \centering
    \includegraphics[width=0.4625\textwidth]{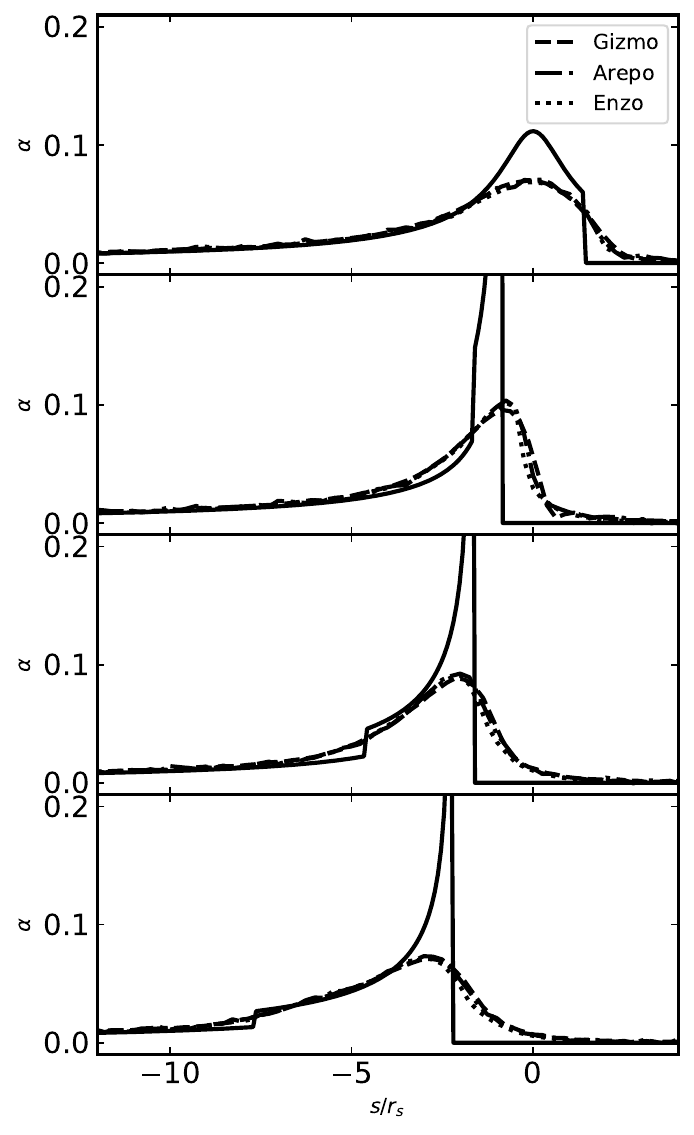}
    \caption{Profile of over-density $\alpha$ along a line of constant $R=2r_s$, at $t=15t_c$, with $\mathcal{M}=0.9,1.1,1.3,1.5$ from top to bottom. Solid lines show the analytic solution, dashed lines show the \texttt{Gizmo} results, dash-dot lines show \texttt{Arepo}, and dotted lines \texttt{Enzo}.}
    \label{fig:theta_line_A0.1}
\end{figure}

\section{Results} \label{sec:results}

In this section, we compare the numerical results from simulations against the corresponding analytic predictions. We focus on the results with the MFM solver, unless otherwise stated. We first compare the structure of the numerical wake with the prediction from linear theory. In particular, we seek to identify the radius beyond which the two agree. This will allow us to quantitatively asses at which radius from an extended perturber dynamical friction can be treated as originating from a point mass. To this end, we compare the net gravitational force acting on the perturber from the perturbed density, across Mach numbers and time.

\subsection{Wake} \label{sec:res_wake}

We directly compare the numerical density distribution with the analytical prediction of the wake form, given by $\alpha$ in Equation (\ref{eq:alpha}). The numerical density is averaged by dividing the particles into cylindrical bins $(s,R)$. This is then normalised to the dimensionless over-density $\alpha=(\rho-\rho_0)/\rho_0$. As a reminder, here $s$ is the distance from the perturber along the direction of travel, and $R$ is the cylindrical radius away from this axis. In our setup, with the perturber fixed and the gas moving past, $s$ is the correct way to refer to the coordinate system centred on the perturber, so that we remain consistent with the Ostriker formulation of the wake. In Figure \ref{fig:wake_zoom}, we show the structure of the numerical over-density, with contours showing the analytic prediction (white dashed lines), for four Mach numbers, $\mathcal{M}=0.9,1.1,1.3,1.5$. The residual (here the difference between numerical and analytic results) shows us that the main difference is in the front of the Mach cone structure. The sharpness of the density peak in the cone has been softened, with the density smeared into a wider profile. The peak in the cone over-density is too small and extends too far forward. We define the approximate initial spatial resolution of the simulation $l$ as the mean particle separation in the initial conditions. The smearing of the profile is on a scale larger than $l$. The profile spread over approximately $20l$. These features are clearly seen in Figure \ref{fig:theta_line_A0.1}, where we show the over-density $\alpha$ on a line of constant $R=2r_s$, for $A=0.1$. Each Mach number is shown in a different panel, from top to bottom $\mathcal{M}=0.9,1.1,1.3,1.5$, with the solid line showing the analytic solution, and the dashed line showing the numerical result. We also compare these MFM solver results to those produced using the MUSCL and PPM schemes.

The spread in the numerical result is caused by the extended nature of the perturber. The analytic solution assumes a point mass. If that mass is instead spread across some spatial distribution, the resultant wake will be approximately a convolution of the point mass solution with the mass distribution. Our results are consistent with previous work, which finds that extended perturbers with softening scale $r_s$ produce spreading from the cone front of $\sim 2r_s$ \citep{ja:furlanetto2002}. The structure is generally well matched at large radii from the perturber, where the over-density is very small. While the exact structure of the wake is important for recovering the perturbation of the background medium, it is the resultant retarding force on the perturber that is often studied.

\subsection{Force} \label{sec:res_force}
We now compare the drag force from the Plummer perturber in our numerical setup to that from the point mass prediction. We consider the force produced across different Mach numbers, and across different parts of the wake. We also show how these forces vary for different numerical resolutions, and include the forces produced by the runs with the alternative solvers. The wake does not match at small radii, as discussed above. However, we aim to identify the radius at which the net force on the perturber matches for both our numerical results, and the point mass prediction from \citet{ja:ostriker1999}. We calculate the force for different inner radial limits $r_\mathrm{cut}$. The force comparison is made to Equation \ref{eq:force}, substituting $r_\mathrm{min}=r_\mathrm{cut}$ so that we are comparing the numerical and analytic forces from the same regions of the respective wakes.


The prediction from linear theory for the net gravitational force on the perturber is given by Equation (\ref{eq:force}). In the subsonic case, this solution to the integral holds for $r_\mathrm{min}<(c_s - V_0)t$, which requires the wake to be larger than the effective size of the perturber. For supersonic cases, the solution is found by assuming $r_\mathrm{min} < (V_0-c_s)t$, which means that the lower limit of the integral is within the Mach cone. A numerical integration of the analytic over-density $\alpha$ is used at times when these conditions were not satisfied. For example, this would happen if the cone is entirely inside $r_\mathrm{min}$, but not the spherical part. This allows us to produce an accurate force for the analytic wake solution, no matter the limits of the integration. It is achieved by explicitly breaking the wake into zones, with cone and sphere sections, and then integrating the sections separately based on where $r_\mathrm{min}$ falls.

The equivalent force from the simulation output is calculated by direct force summation. The Newtonian gravitational force is calculated between each particle of mass $M_\mathrm{part}$ and the massive perturber $M_\mathrm{p}$. All particles within $r<r_{\mathrm{cut}}$ of the perturber are excluded from the calculation.

\begin{figure}
    \centering
    \includegraphics[width=0.45\textwidth]{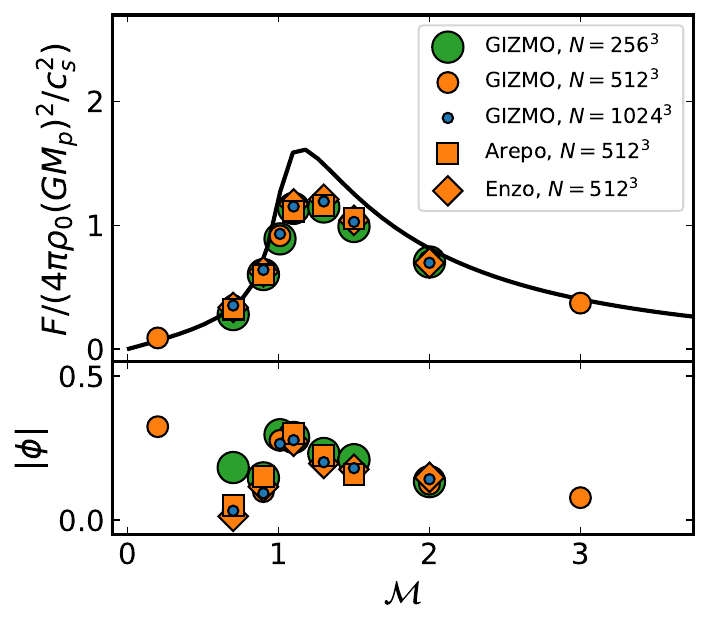}
    \caption{\textit{Upper Panel}: Dimensionless force from numerically induced wakes across a range of Mach numbers, at $t=15 t_c$. We show results from \texttt{Gizmo} (circles), \texttt{Arepo} (squares), \texttt{Enzo} (diamonds), and the corresponding analytic prediction (black line).  We show runs using $N=256^3$ (green), $512^3$ (orange), and $1024^3$ (blue) particles/cells. Both point mass and numerical results forces are calculated using $r_\mathrm{cut}=r_s$. It diverges significantly for supersonic cases close to $\mathcal{M}=1$. The $512^3$ and $1024^3$ agree well, with differences within the 5\% level, while the $256^3$ broadly agree, though do show increased scatter. \textit{Lower Panel}: Absolute residual between the numerical and analytic results $|\phi| = |F_{\mathrm{num}} - F_{\mathrm{ana}}| / F_{\mathrm{ana}}$.}
    \label{fig:force_mach}
\end{figure}

The force is dependent on the Mach number and time. In Figure \ref{fig:force_mach}, we show the variation in the numerical and analytic forces with Mach number for three sets of runs at $t=15t_c$. The three cases are for different resolutions, defined by the number of particles used. We again include results produced with \texttt{Arepo}, as a check against another widely used code. \texttt{Gizmo} results are shown as circles, \texttt{Arepo} as squares, with the colour showing the resolution (see figure for specifics). The bottom panel shows the absolute value of the residual $|\phi|=|F_\mathrm{num}-F_\mathrm{ana}|/F_\mathrm{ana}$. The subsonic setups show reasonable agreement at $\mathcal{M}=0.7$, with some divergence at $\mathcal{M}=0.9$. The residual is significantly higher at $\mathcal{M}=0.2$, due to a very small absolute value of the force. DF at these low Mach numbers is not consequential to actual physical scenarios.

We start by exploring how the numerical Plummer and analytic point mass forces compare when integrating out from the natural inner limit of $r_\mathrm{cut} = r_\mathrm{s}$. The supersonic cases show the most marked differences. The forces are $10 - 25\%$ lower than expected. The difference is most extreme in the supersonic regime close to $\mathcal{M}=1$. At larger Mach numbers $\mathcal{M}\geq 2$, the numerical results show significantly less difference and therefore match better with the point mass solution. The analytic force is also smaller in this regime, so perturbing objects will experience less drag. We therefore focus our analysis and discussion to $0.7 < \mathcal{M} < 2$, where the absolute and proportional differences are most pronounced, and most consequential. This is also the most likely range of Mach numbers for perturbers, such as galaxies and subhaloes orbiting within a host halo. These differences are significant because they show that the solvers used here predict that an extended perturber (in this case a  Plummer perturber) will produce less drag force than if one assumes the point mass \cite{ja:ostriker1999} solution. The latter is commonly used as a subgrid model.

The net drag forces show only small variation with resolution. The green dots in Figure \ref{fig:force_mach} show the runs using $N=256^3$, while the orange dots show $N=512^3$, and the blue dots $N=1024^3$. The orange and blue points match within $5\%$. The green, being the lowest resolution, are more scattered from the underlying pattern, but are still broadly consistent. This demonstrates that we have converged on consistent numerical results, at the $5\%$ level, in terms of both mass and spatial resolution. The MUSCL \texttt{Arepo} and PPM \texttt{Enzo} results fit closely with those from MFM \texttt{Gizmo}. The results have converged when we increase the mass and spatial resolutions, and do not vary between different solvers in different codes. The wake is still smeared out in comparison to the analytic solution. This must come either from the extended nature of the perturber, or the solver. As the results from the different solvers agree closely, we conclude the difference is purely created by the extended perturber.

\begin{figure}
    \centering
    \includegraphics[width=0.45\textwidth]{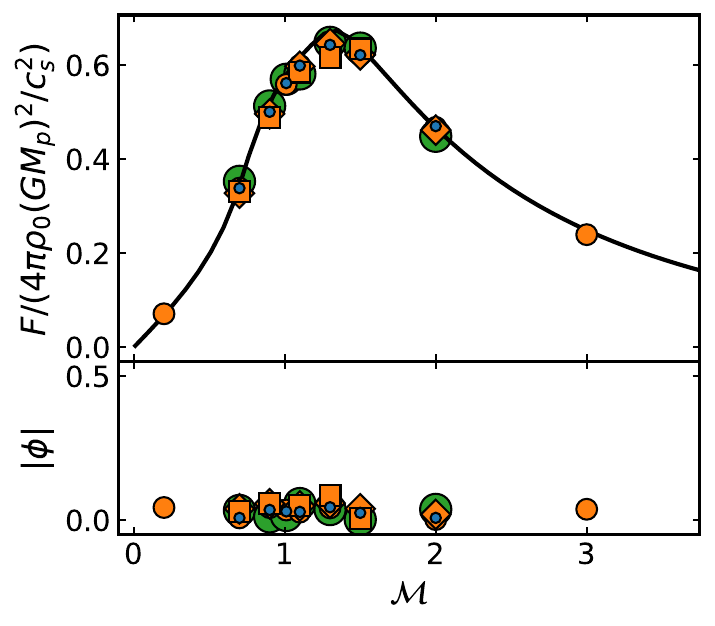}
    \caption{Same as Figure \ref{fig:force_mach}, but using $r_\mathrm{cut}=4r_s$. As before, the different codes/resolutions are given by the coloured symbols (see Figure \ref{fig:force_mach} legend), and the solid black line shows the analytic prediction.} All cases now match within 5\%, in both subsonic and supersonic regimes.
    \label{fig:force_mach_01}
\end{figure}

As presented above, we chose first to calculate results for $r_\mathrm{cut}=r_\mathrm{s}$, as shown in Figure \ref{fig:force_mach}. It is clear that parts of this numerical wake differ from the point mass prediction for the reasons discussed above. However, we are still interested in where the force from an extended perturber is consistent with the equivalent force from a point mass. As one moves away from an extended perturber, the gravitational potential of the perturber appears increasingly like a point mass. To this end we explore at what radius the two become broadly consistent. We find that the force from $r_\mathrm{cut}=4r_\mathrm{s}$ is approximately consistent with the analytic prediction produced for the same inner limit. When $r_\mathrm{cut}=4r_\mathrm{s}$, the numerical force results are within $5\%$ of the expected value. The forces produced when this fit is used are shown in Figure \ref{fig:force_mach_01}. We therefore adopt $r_\mathrm{fit}=4r_s$ as our fit for the radius beyond which the contributions from the numerical and point-mass analytic wakes match one another.

\begin{figure}
    \centering
    \includegraphics[width=0.45\textwidth]{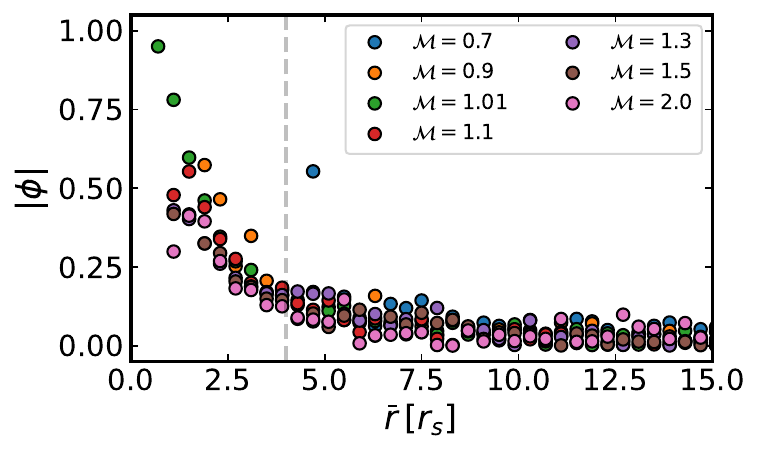}
    \caption{Absolute residual of force contributed from spherical shells of width $r_s$, centred at $\bar{r}$, using $N=1024^3$ runs. We show $\mathcal{M}=0.7$ (blue), $0.9$ (orange), $1.01$ (green), $1.1$ (red), $1.3$ (purple), $1.5$ (brown), $2$ (pink). Dashed grey line marks position of the fit at $r_\mathrm{fit}=4r_s$. Force in all bins beyond this point matches to within $10\%$, at all Mach numbers except $\mathcal{M}=0.7$. Here the values are very small, so numerical variation is high.}
    \label{fig:force_delta_rdr}
\end{figure}

The value for $r_{\mathrm{cut}}$ was derived using the contribution to the force from the wake outside that radius. Since this is an integrated constraint it is interesting to see whether the contribution from radial bins to the force is in agreement as well. Figure \ref{fig:force_delta_rdr} shows a breakdown of the absolute residual, $|\phi| = |F_{\mathrm{num}} - F_{\mathrm{ana}}| / F_{\mathrm{ana}}$,  between the numerical and analytic forces, contributed from spherical shells of width $\sim r_s$ centred at radius $\bar{r}$, for different Mach numbers. We show $\mathcal{M}=0.7$ (blue), $0.9$ (orange), $1.01$ (green), $1.1$ (red), $1.3$ (purple), $1.5$ (brown), $2$ (pink). As before, the largest variation is found close to $\mathcal{M}=1$, with the bulk of the divergence coming from inside $2r_s$. We see that the forces match within $10\%$ per bin for all cases when $\bar{r}\geq4r_s $, except at $\mathcal{M}=0.7$, where the force contributions are all so small that numerical noise creates large variations. Although the results per bin differ at the $10\%$ level, the force of the whole wake is within $5\%$ across our Mach number range. These differences are consistent with the spreading of the sharp cone front, as the impact of this on the gravitational force experienced by the perturber would diminish with distance. Higher Mach cases have more extended cones, therefore the difference is most pronounced very close to the perturber, where the spreading would push some mass in front of the perturber, cancelling some of the drag force. Further out, the spread would all be behind the perturber.

\subsection{Long Term Evolution} \label{sec:res_time}

\begin{figure}
    \centering
    \includegraphics[width=0.45\textwidth]{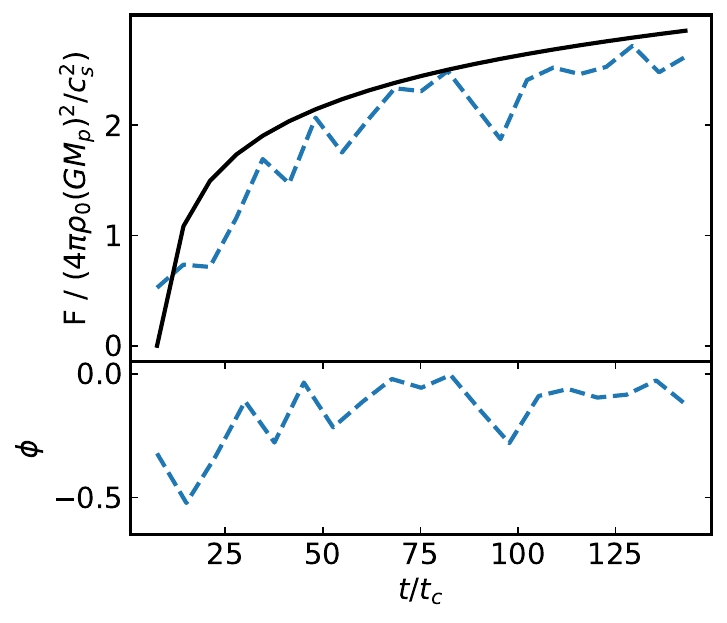}
    \caption{\textit{Upper Panel}: Time evolution of the dimensionless force in the larger L=100kpc box. The crossing time $t_c$ is the sound speed crossing time of the softening scale $r_s$. Blue Dashed line shows force from GM1024M13L, and the solid black line shows the analytic prediction. The force does not converge to the analytic solution in this longer time. \textit{Lower Panel}: Residual between analytic and numerical results $\phi=(F_{\mathrm{num}}-F_{\mathrm{ana}})/F_{\mathrm{ana}}$. Lower integral limit is $r_\mathrm{cut} = r_\mathrm{s}$.}
    \label{fig:force_time_long}
\end{figure}

So far we have shown results $15t_c$ after the perturber is turned on. Some previous works  report results at many hundreds of crossing times \citep{ja:sanchez1999,ja:kim2009}. At these times the wake has reached scales hundreds of times the gravitational softening scale of the perturber. We need to include the whole wake in our analysis, so we are only able to run until the wake reaches the edge of the box. Since the wake grows with $V_0+c_s$, we can only match these much larger numbers of crossing times by using a larger box. We increase the box size by an order of magnitude to achueve this, going from a box edge of $L=10$ to $100$. The specific units are unimportant, due to the scale free nature of the problem, just the relative change matters here. The results of the larger box are shown in Figure \ref{fig:force_time_long}, using GM1024M13L, which extends the run time to $150t_c$. All forces shown here are calculated using $r_\mathrm{cut} = r_\mathrm{s}$. The intrinsic variation in the force is larger for these larger boxes, where $M_\mathrm{part}/M_\mathrm{p} = 8\times10^{-7}$ (see Appendix \ref{app:var}). In summary of this variation, the discretisation of the gas by mass leads to some fluctuation in the net force on the perturber. This fluctuation is greater when the mass ratio of the particle to perturber is large, since a single particle can exert a greater force on the perturber. This effectively places some limits on the setups that can be feasibly run, since very large random variation will prevent meaningful results from being extracted from the simulation output. However, we can still conclude from this longer run that the difference in force from the extended perturber continues to large times.

\subsection{Variation with Resolution}
 
\begin{figure}
    \centering
    \includegraphics[width=0.45\textwidth]{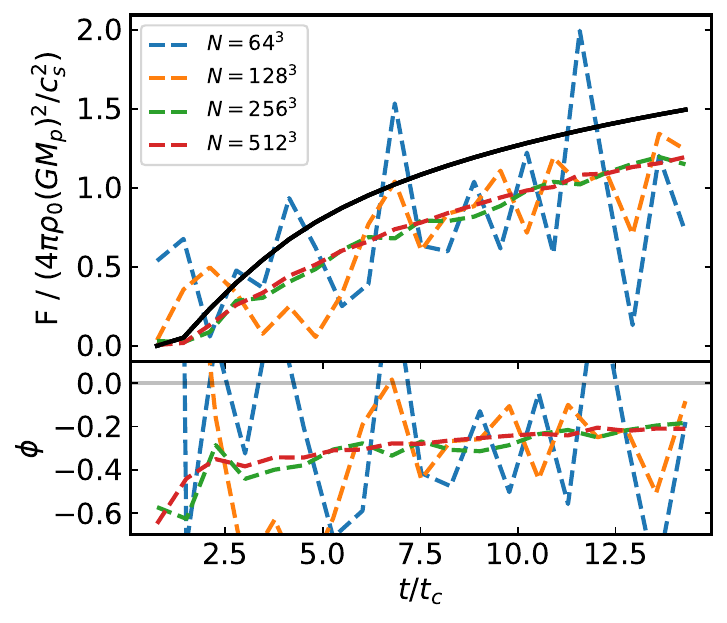}
    \caption{\textit{Upper panel}: Evolution of the dimensionless force for different particle numbers, showing $N=64^3$ (blue), $N=128^3$ (orange), $N=256^3$ (green), $N=512^3$ (red). The solid black line shows the analytic prediction. The evolution is essentially identical for all resolutions, but with significantly increased random variation for $N=64^3$ and $N=128^3$. All forces are calculated using $r_\mathrm{cut} = r_\mathrm{s}$. \textit{Lower panel}: Residual between numerical and analytic wakes $\phi=(F_{\mathrm{num}} - F_{\mathrm{ana}}) / F_{\mathrm{ana}}$.}
    \label{fig:force_time_res}
\end{figure}

Testing the convergence of results based on resolution, via the number of particles, produces the same evolution whether we use $N=64^3$, $128^3$, $256^3$, $512^3$ or $1024^3$ particles, in effect increasing the spatial resolution from $0.8r_s$ to $0.05r_s$. The bulk velocity and sound speed of these runs are identical, with only the particle mass changing. This maintains the same background density. This force evolution is shown in Figure \ref{fig:force_time_res}. The random variation discussed above is effectively dependent on the ratio of the particle mass to the perturber mass. While the lower resolution runs show much greater scatter over time, they are broadly consistent with the higher resolution cases. The exact form of the wake also converges, showing the same structure at $N=256^3$ as at higher resolution. This corresponds to convergence at a spatial resolution of $0.2r_s$. This is roughly equivalent to 5 cells per gravitational softening scale in a grid based code. At resolutions below this, the resolution is too low to practically assess the shape of the detailed structure of the wake.

On top of the tests we present above, we also run verification tests that vary the exact conditions used in the numerical runs, such as perturber mass and initial background gas conditions. These should not impact the results found here, as the problem is effectively scale-free to these variables, and indeed this is what we find. These changes have no impact on the wake structure or the resultant drag force.

The results presented in Sections \ref{sec:res_force} to \ref{sec:res_time} show that the numerical force fits the force predicted for a point masses, when $r_\mathrm{cut}=4r_s$. There is no obvious change in the force trend, even at large times. Our runs have converged, within $5\%$, at the high resolution end, and show good agreement at lower resolutions.

\subsection{Minimum Radius to Produce Force from Whole Wake} \label{sec:res_min_rad}

It is common practice in the analysis of numerical dynamical friction studies to identify a minimum radius $r_{\rm{min}}$ in the analytic point mass solution for which the \citet{ja:ostriker1999} force matches the total drag force found in numerical solutions \citep{ja:sanchez1999, ja:kim2009, ja:bernal2013}. By total drag force, we refer to the force found by integrating the numerical results from $r_\mathrm{cut}=0$ to the edge of the box. Defining such a minimum radius serves two main purposes, it allows to compare different numerical simulations directly with each other and also gives an estimate by how much sub-grid models using the point-mass solution need to be modified for extended perturber. We briefly outline these previous results, and then compare the results produced by our simulations.

\citet[SB99 from here]{ja:sanchez1999} find that using $r_\mathrm{min}=2.25r_\mathrm{s}$ matches their force. \citet[BS13 from here]{ja:bernal2013} estimate the value of $r_\mathrm{min}$ using the position of the maximum density in the wake produced by a Plummer sphere, as opposed to the position of a point mass, finding a value consistent with the SB99 result. \citet[KK09 from here]{ja:kim2009}, on the other hand, find a Mach dependent prescription, with $r_\mathrm{min} = 0.35 \mathcal{M}^{0.6} r_\mathrm{s}$ and argue that the difference between their result and previous ones is caused by differences in resolution and equation of state. BS13 address this suggestion, and find no difference in results between isothermal and adiabatic equations of state, or with resolution. They use the $r_\mathrm{min}=2.25r_\mathrm{s}$ value, but also explore a Mach dependent fit, resulting in $r_\mathrm{min} = 1.5 \mathcal{M}^{0.6} r_\mathrm{s}$. BS13 conclude that it is only applicable at later times $t > 130t_c$ for supersonic perturbers, where $t_c$ is the background medium's sound speed crossing time of the gravitational softening scale $r_s$. Both fits discussed by BS13 use a value significantly larger than that found by KK09, which has a dramatic effect on the absolute value of the drag force from the wake. BS13 note that the KK09 fit can produce a force as much as a factor of two larger than their own.

\begin{figure}
    \centering
    \includegraphics[width=0.45\textwidth]{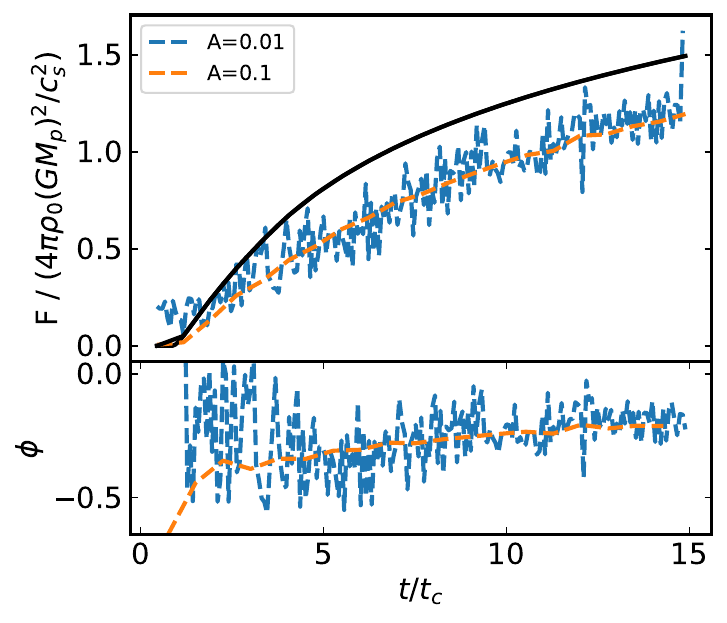}
    \caption{\textit{Upper panel}: Evolution of the dimensionless force for $A = 0.01$ (blue) and $A = 0.1$ (orange), using runs GM512M13LM and GM512M13 respectively. The evolution is essentially identical for all resolutions, but with significantly increased random variation for $N=64^3$ and $N=128^3$. All forces are calculated from $r_\mathrm{cut}=r_\mathrm{s}$ outwards. The solid black line shows the analytic point mass solution for $r_\mathrm{min} = r_\mathrm{s}$. \textit{Lower panel}: Residual between numerical and analytic wakes $\phi=(F_{\mathrm{num}} - F_{\mathrm{ana}}) / F_{\mathrm{ana}}$.}
    \label{fig:force_time_a_comp}
\end{figure}

The previous studies use runs at $A=0.01$. While most of our runs use $A=0.1$, we have run a comparison case using $A=0.01$ GM512M13LM, achieved by reducing the perturber mass by an order of magnitude, and which does not show a different evolution. This case shows a much greater innate variance due to the lower ratio of particle mass to perturber mass (see Appendix \ref{app:var} for a detailed explanation). The overall trend in force from this case, however, is the same as the $A=0.1$ case, but with increased variance. This is shown in Figure \ref{fig:force_time_a_comp}, with $A=0.01$ in blue and the standard $A=0.1$ in orange.

We find $r_{\mathrm{min}}=0.8r_s$ which implies that our total drag force (integrating from $r_s$ outwards) is $20\%$ smaller than the total drag force reported in KK09, but $60\%$ larger than that from BS13. We carry out this calculation for $\mathcal{M}=1.3$, where we have the most comprehensive data. A future study could conduct a more comprehensive comparison to the differences in the $r_\mathrm{min}$ values found by the various studies.

\section{Discussion} \label{sec:discussion}

This work represents the first numerical study of dynamical friction using this class of meshless hydro solver, in this highly idealised context. We find a fit of $r_\mathrm{fit}=4r_s$ beyond which the wake produces the retarding force predicted by linear theory for a point mass perturber. This value is somewhat different from previous numerical studies, but broadly consistent. Below we discuss implications for cosmological simulations, and the usefulness of this setup as a standard test for hydrodynamics modelling in astrophysics.

\subsection{Implications for Cosmological Simulations}
We have shown that the force only matches the linear theory beyond our $r_\mathrm{fit}$, across many crossing times, using the MFM, PPM and MUSCL-scheme solvers. Here we discuss how the idealised setup used here could map onto the dark-matter substructure in cosmological simulations, and if we can draw any conclusions about the capturing of the DF they should experience.

\begin{figure}
    \centering
    \includegraphics[width=0.45\textwidth]{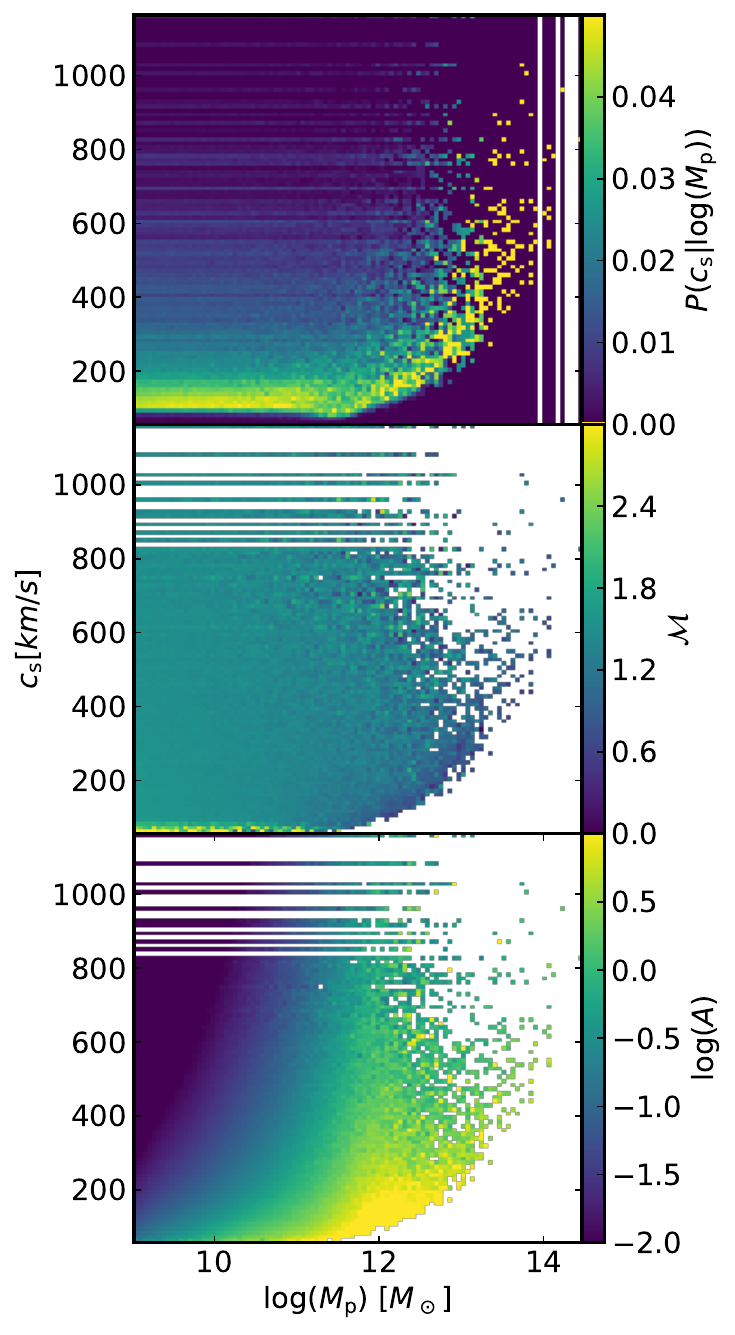}
    \caption{Subhaloes from the IllustrisTNG-300 simulation box at $z=0$. Sound speeds are calculated from the virial temperature of the host halo, dependent on the virial mass $M_{200}$. All panels show the subhaloes with masses between $M_\mathrm{p}=1\times10^{11} M_\odot$ and $1\times10^{15}M_\odot$, binned into subhalo mass - sound speed pixels. \textit{Top panel}: The colour in this panel indicates the conditional probability of finding a sound speed, given a certain subhalo mass. \textit{Middle panel}: Average Mach number of the subhaloes in each $(M_\mathrm{p},c_\mathrm{s})$ pixel. The subhalo velocity is taken as the peculiar velocity of the subhalo, relative to the velocity of the host halo. \textit{Bottom panel}: Average $A$ value in each $(M_\mathrm{p},c_\mathrm{s})$ pixel. The softening scale equivalent radius is taken to be the radius of the max velocity, based on the density profile of the subhalo.}
    \label{fig:MpCsAll}
\end{figure}

\begin{figure}
    \centering
    \includegraphics[width=0.45\textwidth]{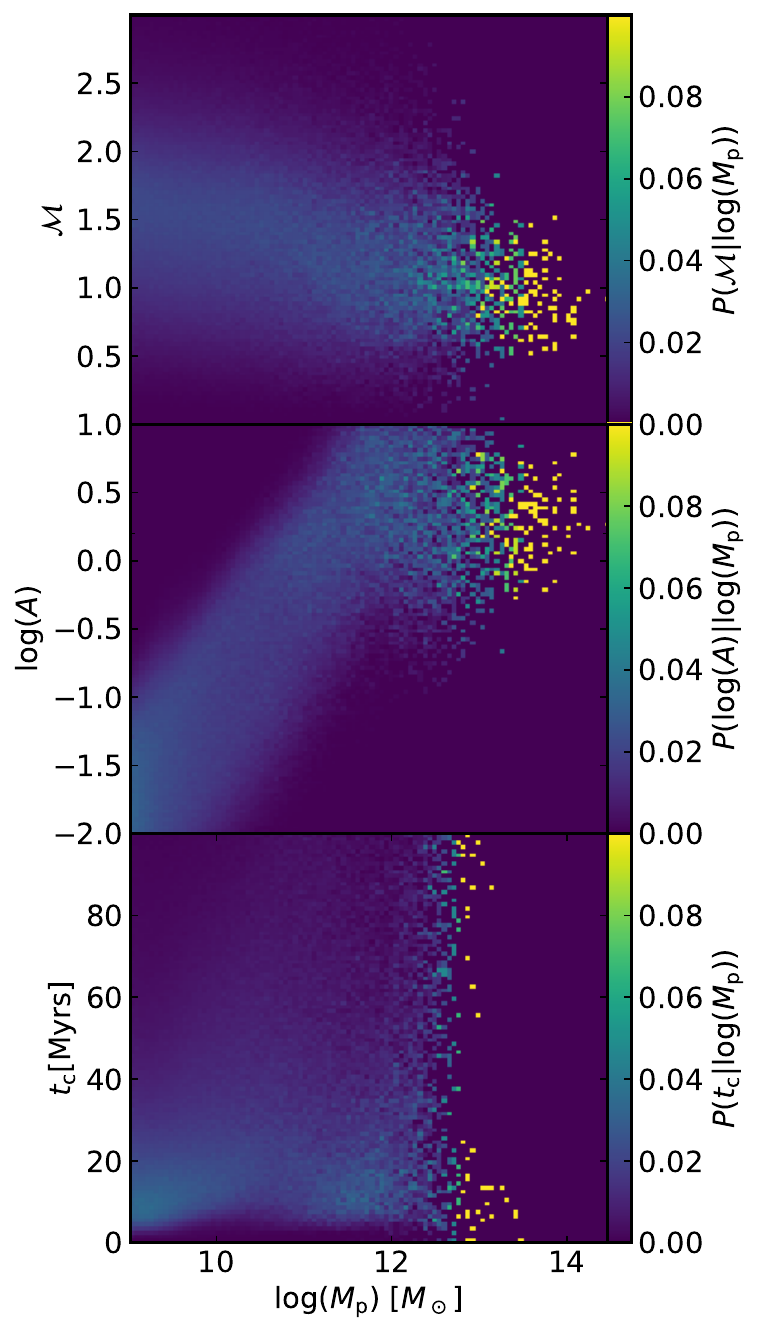}
    \caption{\textit{Top panel}: Conditional probability of finding a Mach number, given a subhalo mass. The majority of subhaloes have mach numbers in the $\mathcal{M}=1-2$ range. \textit{Middle panel}: Conditional probability of finding an $A$ number, given a subhalo mass. subhaloes with masses below $10^{10}M_\odot$ all exist well within the linear regime ($A<<1$). Above this mass, a significant fraction in each mass bin are either in the transition regime, or into the non-linear regime. \textit{Bottom panel}: Conditional probability of finding a crossing time, given a subhalo mass. This crossing time is defined as the sound speed crossing time of the radial extent of the subhalo. For subhaloes of mass $10^9M_\odot$ to $10^{12}M_\odot$ there is large spread of crossing times, with broad peak at $t_\mathrm{c}=10-30$ Myrs.}
    \label{fig:MachAHist}
\end{figure}

The typical conditions in which we find subhaloes in cosmological simulations can be mapped onto the  idealised point mass setup by estimating their approximate Mach number and $A$ parameters. We have taken the simulation results from the IllustrisTNG-300 public data \citep{ja:nelson2018}, using the halo and subhalo catalogues at $z=0$. We selected all host haloes in the mass range $M_p=10^{11}M_\odot$ to $10^{15}M_\odot$. The distribution of masses and sound speeds for all subhaloes are shown in Figure \ref{fig:MpCsAll}. The top plot shows the conditional probability of a sound speed, given a subhalo mass. The sound speed depends only on the mass of the host (see Appendix \ref{app:sims}), which leads to the bands at the end of high sound speed/host mass, where there are the fewest samples. We effectively see the negative slope in the power law distribution of halo/subhalo masses, as there are many more subhaloes associated with low mass haloes (low sound speeds), even though the most massive haloes individually contain far more subhaloes than any given lower mass halo. The middle plot shows the mean Mach number for each pixel. The distribution is fairly uniform across both mass and sound speed. The mean Mach number is around $\mathcal{M}=1-2$. This is also shown in the top panel of Figure \ref{fig:MachAHist} with the conditional probability of a Mach number, given a subhalo mass. The bottom plot of Figure \ref{fig:MpCsAll} shows the mean A parameter for each pixel. The distribution at the high subhalo mass end is dominated by small numbers of very high $A$ parameter values in each cell. The overall distribution of $A$ parameters are shown in the middle panel of Figure \ref{fig:MachAHist}. The distribution has been truncated to $A=10$, but continues to values in the thousands. We show the conditional probability of finding an $A$ number, given a subhalo mass. The subhaloes below $M_p=10^{10}M_\odot$ exist well within the linear regime ($A<<1$). Those between $M_p=10^{10}M_\odot$ and $10^{11}M_\odot$ show a wider range of $A$, numbers, but still mainly reside in the linear or quasi-linear regime ($A\leq1$). Above this mass, the distribution spreads significantly, with a range of $A$ numbers, a significant fraction in the non-linear regime ($A>1$).

The IllustrisTNG-300 subhaloes show that a large fraction of subhaloes in a typical cosmological simulation exist within the linear DF regime, with $A<1$, and with Mach numbers in the range $\mathcal{M}=1-2$. This is the regime in which we have performed our numerical study, and the one in which gaseous DF is most impactful. The bottom panel of Figure \ref{fig:MachAHist} show the distribution of sound speed crossing times for the radial extents of these subhaloes. We see that these values are of the order $10-100$ Myrs. 

A number of factors make DF in a system with more complete physical processes diverge from the analytic solution. The gas will experience self-gravity, leading to the further growth of the overdense wake. (To be clear, in the idealised cases presented earlier the gas does not experience self-gravity, but in the full physics simulation, this would be present.) Radiative cooling mechanisms move the gas away from the adiabatic scenario considered in the analytic solution. Feedback mechanisms from star formation, such as supernova driven winds, will also have complex effects. These will all be present in the full cosmological context of a cosmological simulations. The effect of these processes will be to significantly change the structure of the wake in complex ways.

While we have assumed, in our idealised setup, that the background medium has a uniform density, the simulated medium through which the subhaloes move will be far from uniform. The CGM is highly variable, made up of gas with a huge range of sound speeds. The wake will not build up smoothly as shown in the idealised scenario, where it can grow for an unlimited time through an infinite uniform background medium. Instead, at any given time, the wake will be built from the medium that the perturber (subhalo) is moving through at that time.

All of these processes aside, one major aspect of the environments around these substructures is how many particles make up the wake, however it ends up structured. Our study includes runs with as few as $N=64^3$ particles (see Figure \ref{fig:force_time_res}), but this is likely still larger than the numbers of particles used to resolve the volume around a given substructure. We found that the force from the wake at this resolution shows large random variations, created by numerical noise, but that it does vary around the same value as that found at much higher resolution. It is therefore plausible that DF over longer enough time scales, is being captured correctly in these simulations. This is all the more important because many of these subhaloes exist at high redshift, where gas fractions are higher, and thus gaseous DF has its strongest effect.

\subsection{Gaseous DF as a Standard Grav-Hydro Test}
As we have seen in the preceding sections, the idealised gaseous dynamical friction problem presents a tough challenge for modern gravo-hydrodynamical solvers. It combines both gravitational forces and fluid dynamics, and includes complex flows which will not be aligned with any underlying structured mesh. The presence of the Mach cone, with its sharp density profile, one which is not aligned directly with, or orthogonal to, the bulk flow is a particular challenge These factors make this setup an ideal candidate for a  standard numerical test case for multi-physics astronomical codes. The problem has a well-defined analytic prediction for both the structure of the overdense wake, and the force produced by that wake in the linear regime. This allows for direct assessment of the accuracy of the numerical results, something that is relatively rare for tests that include complex, multidimensional fluid flows, one which includes the effects of both fluid and gravitational forces. The process of DF is also present across a range of important astrophysical problems and is a critical part of many of them, so understanding the abilities of our numerical methods in this area is important for our wider understanding of the underlying physics of these systems.

An additional strength of this problem, as a standardised test, is its scale-free nature. Any setup of the problem can be exactly defined by only two parameters, the $A$ value, which effectively defines the linearity of the problem, and the Mach number $\mathcal{M}$. Any combination of sound speed, perturber mass, and potential softening can be compared using these scale-free parameters. The only limit in comparing to the analytic solution is that of linearity. In other words, the analytic solution assumes the wake is linear at all positions used in the comparison, such that the comparison is best made at $A<1$. It is therefore extremely easy to directly compare multiple sets of numerical results to the analytic prediction, and to one another. For instance, reproducing plots such as that shown in Figure \ref{fig:force_mach}, compares results across $A$ and $\mathcal{M}$ to the analytic solution.

We propose this setup of a massive perturber, represented by some softened fixed potential, embedded within a gaseous medium moving with some bulk flow, be used as a standard test for current and new multi-physics codes. It concisely assesses the accuracy of both hydrodynamic and gravity solvers, used in conjunction, with a well defined solution. In particular, using results close to $\mathcal{M}=1$, where the gaseous response is most pronounced, and the drag force is greatest, will produce interesting comparisons.

\section{Conclusions}
We present a suite of idealised gaseous dynamical friction simulations, using a class of Lagrangian hydrodynamical solver widely used in astrophysics, but not before tested in this manner. Specifically, the MFM solver, in \texttt{Gizmo}, with comparisons to runs using \texttt{Gadget4}, \texttt{Arepo} and \texttt{Enzo}. We find that the structure of the gravitationally induced wakes are broadly consistent with previous numerical works, and with the predictions of linear theory for a point mass perturber, once the effects of the extended perturber are accounted for. In summary, we find that:
\begin{itemize} 
    \item the drag force from the overdense wake of our extended perturber fits the linear prediction for a point mass beyond $r_\mathrm{fit}=4r_s$, to within $5\%$
    \item we have converged with resolution, from as few as 5 cells per $r_s$, as well as in time, such that these results should be applicable to any equivalent setup using this solver
    \item the results for the different solvers are in agreement, be it with the MUSCL scheme in \texttt{Arepo}, or the PPM solver in \texttt{Enzo} (both of which codes are public)
    \item all these conclusions are independent of choices made in our exact setup, be it physical conditions (e.g background density/temperature) or numerical parameters (e.g time-step).
\end{itemize}

It is clear that DF is a complex problem for modern hydro solvers to treat. DF is a process that is present in astrophysical systems across a host of mass and length scales and is crucial to our understanding of the evolution of these systems.

\section*{Acknowledgments}

The authors would like to thank S. Gordon for many useful discussions, and their invaluable help with \texttt{Enzo}.

This work was performed using the Cambridge Service for Data Driven Discovery (CSD3), part of which is operated by the University of Cambridge Research Computing on behalf of the STFC DiRAC HPC Facility (www.dirac.ac.uk). The DiRAC component of CSD3 was funded by BEIS capital funding via STFC capital grants ST/P002307/1 and ST/R002452/1 and STFC operations grant ST/R00689X/1. DiRAC is part of the National e-Infrastructure.

We acknowledge that the results of this research have been achieved using the DECI resource ARCHER based in the UK at EPCC with support from the PRACE aisbl.

This work also used the Cirrus UK National Tier-2 HPC Service at EPCC (http://www.cirrus.ac.uk) funded by the University of Edinburgh and EPSRC (EP/P020267/1).


The author thankfully acknowledges RES resources provided by University of Valencia in Tirant to RES-AECT-2023-2-0026.

JO acknowledges support from grants PID2022-138855NB-C32 and CNS2022-135878 funded by MICIU/AEI/10.13039/501100011033 and European Union NextGenerationEU/PRTR.

For the purpose of open access, the author has applied a Creative Commons Attribution (CC BY) licence to any Author Accepted Manuscript version arising from this submission.

\section*{Data Availability}

All data produced from the simulations discussed in this work are available on request.




\bibliographystyle{mnras}
\bibliography{df} 



\appendix
\section{Intrinsic Force Variation} \label{app:var}

\begin{figure}
    \centering
    \includegraphics[width=0.45\textwidth]{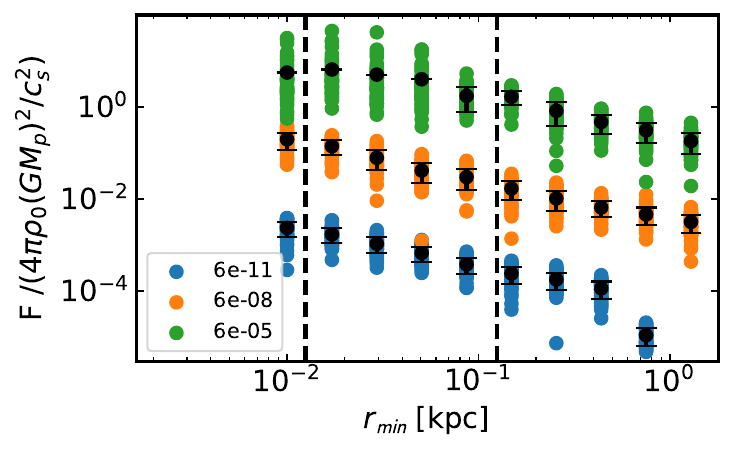}
    \caption{Force on a $M_\mathrm{p}=2.5x10^3 M_\odot$ perturber from the initial particle distribution for setups with mass ratios of $M_\mathrm{particle}/M_\mathrm{perturber} = 6\times10^{-11}$ (light blue dots), $6\times10^{-8}$ (orange dots), $6\times10^{-5}$ (green dots). ICs are randomly shifted 50 times for each tested $r_{\mathrm{min}}$. The coloured dots show the net dimensionless force on the perturber at the centre of the box, excluding all particles within $r_{\mathrm{min}}$ of the perturber. The black dots show the mean force at that $r_{\mathrm{min}}$, with error bars showing the standard deviation in the forces. The variation rises significantly with particle-perturber mass ratio, and falls as $r_\mathrm{min}$ is increased. The variation shown here effectively limits the scenarios that can be run with these solvers.}
    \label{fig:random_force}
\end{figure}

The $A$ parameter, which effectively sets the linearity of the numerical setup, can be varied by changing any of its dependent variables: the perturber mass, the sound speed of the background gas, and the gravitational softening scale of the perturber. Setups with the same $A$ parameter constructed with different combinations of these parameters should produce the same results. However, there is an intrinsic error in the force produced by a given setup, effectively dictated by the inner radius, which forms the lower limit of the force integral, the mass of the perturber, and the mass of the gas particles. If we consider a scenario where the perturber does not act on the medium, then as the initial distribution of particles moves past the perturber position, the configuration of finite masses will produce variation in the net force on the perturber. This force from the unperturbed medium should be zero, but the sampling of the medium by finite mass particles leads to some variation from this value. The lower limit of the integral sets the distance of the nearest particle included in the force summation. Small lower limits lead to large contributions from the particles very close to the perturber, where the small volumes are sampled by few particles. Large $r_\mathrm{min}$ leads to smaller variation. For a given background density and integral lower limit, the ratio of the particle mass to perturber mass $M_\mathrm{part}/M_\mathrm{p}$ has a significant role in determining the level of intrinsic variation. The larger this ratio is, the more severe the impact of the closest particles to the perturber. Figure \ref{fig:random_force} shows the force on the central massive perturber from the glass like initial conditions. The force was calculated by direct summation of the Newtonian gravitational force between the central mass and the gas particles, excluding particles within $r_\mathrm{min}$ of the centre. This effectively shows the uncertainty in the force for different choices of the inner radius. The specific cases shown here are for the $M_\mathrm{p}=2.5x10^3 M_\odot$ perturber and $N=512^3$ particles, with $L=1$, $10$, $100$  $\mathrm{kpc}$. The ratios of particle mass to perturber mass are $6\times10^{-11}$ (light blue), $6\times10^{-11}$ (orange), and $6\times10^{-5}$ (green) respectively.

The numerical dimensionless force increases linearly with the ratio of particle mass to perturber mass, for a given background density. We can estimate the underlying uncertainty in the force from the output of a given setup. This intrinsic variation effectively limits the setups that will produce useful results, because the error can be too large to say anything meaningful about the force from the resultant wake. We must balance the need for a large box to reach large numbers of crossing times with the need to keep the number density high.

\section{Simulation Data} \label{app:sims}

We are using haloes and subhaloes from the IllustrisTNG-300 simulation box \citep{ja:nelson2018}, at redshift $z=0$, to assess the conditions in which simulated subhaloes are found. We select all host haloes in the mass range $10^{11}M_\odot$ to $10^{15}M_\odot$. For each host in this range, we then assign a sound speed. This sound speed is calculated from the virial temperature of that halo
\begin{equation}
    c_s = \sqrt{\frac{\gamma k_B T_{\mathrm{vir}}}{\mu m_p}},
\end{equation}
where $\mu$ is the mean molecular weight of the gas, $m_p$ is the mass of a proton. The virial temperature $T_\mathrm{vir}$ is found by assuming that the gas falling onto the virial radius transforms its kinetic energy into thermal energy. If we assume this produces an isothermal sphere, the virial temperature is given by
\begin{equation}
    T_\mathrm{vir}=\frac{1}{3}\frac{\mu m_p}{k_B}\frac{G M_{200}}{r_{200}}.
\end{equation}
The virial mass is taken as the mass $M_{200}$ enclosed within the radius $r_{200}$. This is the radius at which the the average density of the enclosed material falls to two hundred times the critical density $\rho_\mathrm{crit}$. The subhaloes associated with these hosts are then binned by their mass and this sound speed. The most massive subhalo in each host is excluded, as it is assumed to be the central object of that halo.

These subhaloes are taken as the perturbing massive objects, as they move through the extended gaseous medium of their host. The peculiar velocity of the subhalo, relative to the peculiar velocity of the host halo, defines the velocity for the calculation of the subhalo's Mach number. The radial extent of the subhalo is the equivalent to the softening scale of the Plummer potential. This is taken as the radius at which the density profile of the subhalo produces the maximum circular velocity $V_\mathrm{max}$. The mass of the subhalo, which is the equivalent quantity to the mass of the perturber, is simply the total baryonic and dark matter mass associated to that subhalo by the halo identification algorithm. 


\bsp	
\label{lastpage}
\end{document}